\documentclass[a4paper,11pt,twocolumn]{article}
\usepackage{psfig}
\usepackage{amsmath}
\pssilent
\psfigurepath{figs}
\usepackage{amsmath}
\usepackage{ifthen}
\newcommand{\eq}[2][]{
	\ifthenelse{\equal{#1}{}}{
		\begin{equation*}
			#2
		\end{equation*}
	}{
		\begin{equation}
			\label{#1}
			#2 
		\end{equation}
	}
}
\newcommand{\meq}[2][{}]{
	\ifthenelse{\equal{#1}{}}{
		\begin{equation*}
			\begin{split}
				#2
			\end{split}
		\end{equation*}
	}{
		\begin{equation}
			\label{#1}
			\begin{split}
				#2
			\end{split}
		\end{equation}
	}
}

\title{Phase Transitions in a Probabilistic Cellular Automaton with Two
Absorbing States}

\author{
	F. Bagnoli$^{(1,2)}$,\thanks{
		also INFN and INFM sez. di Firenze;
		e-mail:~bagnoli@dma.unifi.it
	}
 	N. Boccara$^{(2,3)}$\thanks{
 		e-mail: nboccara@amoco.saclay.cea.fr and boccara@uic.edu
 	}
 	and 
 	P. Palmerini$^{4}$\thanks{
 		e-mail: palmerini@dma.unifi.it
 	}
 	\vspace{.5cm}
 	\\
	{
 		\small
 		\begin{minipage}{14cm}
 			\begin{enumerate}
				\item Dipartimento di Matematica Applicata ``G. Sansone'', 
					Universit\`a di Firenze, via S. Marta, 3 I-50139, Firenze, Italy.
				\item DRECAM/SPEC, CE-Saclay, F-91191 Gif-sur-Yvette Cedex, France.
				\item Department of Physics, University of Illinois, Chicago, USA.
				\item Dipartimento di Fisica, Universit\`a di Firenze, 
						Largo E. Fermi 2, I-50125 Firenze, Italy
			\end{enumerate}
		\end{minipage}
	}
	\vspace{.5cm}
	\\
	\begin{minipage}{16cm}
		\begin{center}
			{\bf Abstract}
		\end{center}
	\end{minipage}
	\vspace{1ex}
	\\
	\normalsize
	\begin{minipage}{16cm}
		\noindent We study the phase diagram and the critical behavior of a one-dimen\-sional
		radius-1 two-state totalistic probabilistic cellular automaton having two
		absorbing states. This system exhibits a first-order phase transition between
		the fully occupied state and the empty state, two second-order phase
		transitions between a partially occupied state and either the fully occupied
		state or the empty state, and a second-order  damage-spreading phase
		transition. It is found that all the second-order  phase transitions have the
		same critical behavior as the directed percolation model. The mean-field
		approximation gives a rather good qualitative description of all these phase
		transitions.
	\end{minipage}
}
\date{}
\begin{document}
\maketitle

\section{Introduction}
\label{introduction}
Probabilistic cellular automata (PCA) have been widely used to model a variety
of systems with local interactions in physics, chemistry, biology and social
sciences~\cite{Farmer_etal_1984,Wolfram_1986,Manneville_etal_1989,%
Gutowitz_1990,Boccara_etal_1993}. Many of these models exhibit transcritical
bifurcations which may be viewed as second-order phase transitions. It has been
conjectured~\cite{Jensen_1981,Grassberger_1982} that all second-order phase
transitions from an ``active'' state characterized by a nonzero scalar order
parameter to a nondegenerate ``absorbing'' state provided that (i) the
interactions are short range and translation invariant, (ii) the probability at
which the transition takes place is strictly positive, and (iii) there are no
multicritical points belong to the universality class of directed percolation
(DP). 
In the Domany--Kinzel (DK) cellular automaton~\cite{Domany_Kinzel_1984,%
Kinzel_1985}, which is the  simplest PCA of this type, Martins et.\
al.~\cite{Martins_etal_1991}
discovered a new ``phase.'' These authors considered two initial random
configurations differing only at one site, and studied their evolution with
identical  realizations of the stochastic noise. The site at which the initial
configurations  differ may be regarded as a ``damage'' and the question they
addressed was whether this damage will eventually heal or spread. The new
phase, they called ``chaotic'', is characterized by a nonzero density of
damaged sites. In a recent paper, Grassberger~\cite{Grassberger_1995} (see also
Ref.~\cite{Bagnoli_1996})
showed that this new phase transition also belong to the universality class of
directed percolation. 

\begin{figure}[t]
	\begin{center}
	\hspace{0cm}\psfig{figure=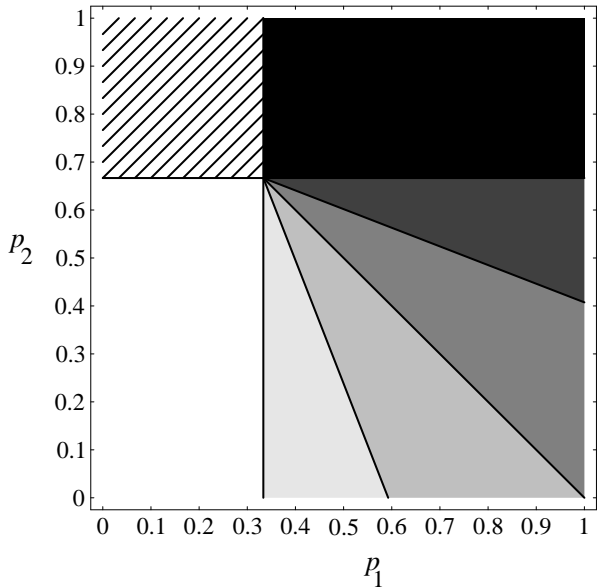,width=8cm}%
	\end{center}
	\caption{Mean-field phase diagram for the density $c$ of active
		sites for $p_3=1$, 
		as described by Equation (1).
		The gray levels indicates the asymptotic value of $c$; contours for 
		$c=0$, $1/4$, $1/2$, $3/4$ and 1. The dashed region marks coexistence
		of phases. The first order phase boundary depends on the initial
		condition.
	}
\end{figure}

More recently, Hinrichsen~\cite{Hinrichsen_1997} stressed that in presence of 
multiple absorbing states with symmetric weight, the universality class
changes from DP to a parity conservation (PC) class, with quite different 
exponents.  

The damage-spreading phase transition can be related to usual chaotic
properties  (i.e.\ positivity of maximal Lyapunov exponent) of dynamical
systems~\cite{Bagnoli_etal_1992}.  The boundary of the damaged phase depend in
general on the algorithm used for the implementation of the evolution rule.
Recently, however, Hinrichsen~\cite{Hinrichsen_etal_1997} formulated an
objective, algorithm-independent definition of damage spreading transitions. In
particular, the algorithm with maximal correlation between the random numbers
used in the updating  (i.e.\ using only one random number) allows the
computation of  the minimum boundary of damage spreading transitions: inside
this region a damage will spread  whichever algorithm is used.  
 
In this paper, we study the phase diagram and the critical behavior of a
one-dimen\-sional radius-1 totalistic PCA having two absorbing states, and
which exhibits a multicritical point. In the DK model, which is a two-input PCA
characterized by the transition probabilities $P(1| 00)=1$, $P(1| 01)=P(1|
10)=p_1$ and $P(1| 11)=p_2$,  the system has two absorbing states for $p_2=1$,
and exhibit a first-order phase transition at $p_1=1/2$. We may therefore,
expect that our system will have a nontrivial phase diagram. Moreover, the
discretization of differential equations lead naturally to radius-1 rules (or
three-inputs rules) and we feel that it would be useful to have a better
knowledge of the phenomenology of simple models formulated in terms of radius-1
PCA.

\section{The model}
\label{themodel}
The evolution rule of our model is defined as follows. Let $s(i,t)$ denotes the
state of the $i$-th cell at time $t$, and $\theta(i,t)=s(i-1,t)+s(i,t)+s(i+1,t)$
 the sum of the cells in the neighborhood, then  
\eq{
	s(i,t+1)=\begin{cases}
		0, 		&\text{if $\theta(i,t)=0$,}\\
		X_1,  &\text{if $\theta(i,t)=1$,}\\
		X_2,  &\text{if $\theta(i,t)=2$,}\\
		X_3,  &\text{if $\theta(i,t)=3$,}
	\end{cases}
}
where $X_j$ $(j=1,2,3)$ is a Bernoulli random variable equal to $1$ with
probability $p_j$, and to $0$ with probability $1-p_j$. That is, the transition
probabilities of this model are $P(0| 000)=1$, $P(1| 001)=P(1| 010)=P(1|
100)=p_1$,  $P(1| 011)=P(1| 101)=P(1| 110)=p_2$, and $P(1| 111)=p_3$.  Cell $i$
is said to be ``empty'' at time $t$ if $s(i,t)=0$, and ``occupied'' if
$s(i,t)=1$. The state in which all cells are empty is a fixed point (absorbing
state) of the dynamics. If $p_3=1$, the state in which all cells are occupied
is also a fixed point.  In this work we study mainly the case $p_3=1$, delaying
the study of the case $p_3<1$ to section 5. 

\begin{figure}[t]
	\begin{center}
	\hspace{0cm}\psfig{figure=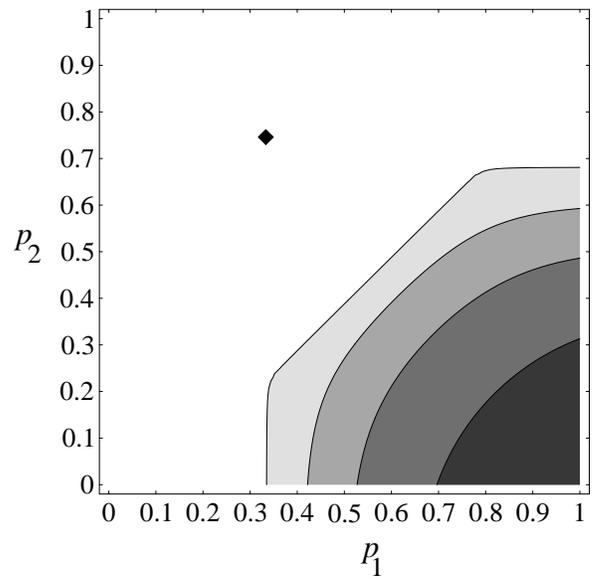,width=8cm}%
	\end{center}
	\caption{Mean-field damage-spreading phase diagram for $p_3=1$.
		 The phase diagram has been obtained numerically iterating
		Equation (2). Levels of damage are $h=0$, $1/8$, $1/4$, $3/8$ and $1/3$.
		The isolated dot marks the multicritical point.
	}
\end{figure}

If there exists a stable ``active'' state such that the asymptotic density $c$
of occupied cells is neither zero nor one, then the model will exhibit various
bifurcations between these different states as the parameters $p_1$ and $p_2$
vary. This totalistic PCA can be viewed as a simple model of opinion formation.
It assumes that our own opinion and the opinion of our nearest neighbors have
equal weights. The role of social pressure is twofold. If there is homogeneity
of opinions, then one does not change his mind (absorbing states), otherwise
one can  agree or disagree with the majority with a certain  probability.

\begin{figure}[t]
	\begin{center}
	\hspace{0cm}\psfig{figure=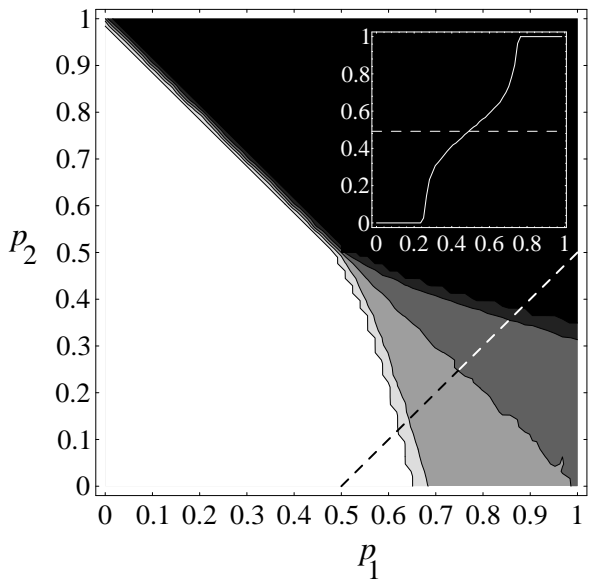,width=8cm}%
	\end{center}
	\caption{Phase diagram for the density of active sites $c$
		 obtained by direct simulations. One run was performed.
		Lattice
		size: 10000, number of time steps: 10000, resolution: 64 different values
		for both
		$p_1$ and $p_2$ ($p_3=1$), color codes as in Figure 1. The
		initial density is $c_0=0.5$.
		The inset represents the density profile along the dashed line. Two critical
		phase transitions are evident.
	}
\end{figure}

Let us assume that one starts with a random configuration half filled with 0's
and 1's. On the line $p_1+p_2=1$, if $p_1$ is small, to first order in $p_1$,
clusters of ones perform symmetric random walks, and have equal probabilities
to shrink or to grow. Slightly off this line, the random walks are no more
symmetric, and according to whether $p_1$ is greater or less than $1-p_2$, all
the cells will eventually be either occupied or empty. $p_1+p_2=1$ is,
therefore, a first-order transition line. This line, however, cannot extend to
$p_1=1$ since, for $p_2=0$, our model is similar to the diluted XOR rule (Rule
90) of the DK model. We thus expect that our model will exhibit, for a certain
critical value of $p_1$, a second-order phase transition coinciding with the
damage-spreading phase transition. For $p_1=1$ and $p_2=0$, frustration is
large, and our model is just the modulo-2 rule (Rule 150) as in the DK model.

\begin{figure}[t]
	\begin{center}
	\hspace{0cm}\psfig{figure=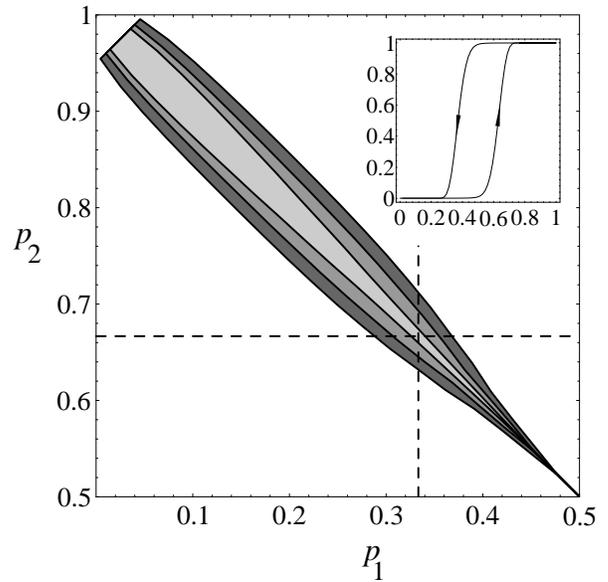,width=8cm}%
	\end{center}
	\caption{
		Profile of the hysteresis region for several values of the noise
		$\varepsilon$ and relaxation time $T$. Data obtained from the local
		structure approximation of order $l=6$; the regions corresponds to the 
		intersections of the hysteresis cycle with $c=0.5$ (horizontal dashed line
		in the inset). The hysteresis boundary lines smoothly join  at $p_1=0$,
		$p_2=1$ (not represented). Darker to lighter areas correspond to $T=500$,
		$\varepsilon=0.0001$; $T=1000$, $\varepsilon=0.0001$; $T=500$,
		$\varepsilon=0.001$. The dashed lines represent the mean-field hysteresis
		region. The inset represents the cycle along a line parallel to the
		diagonal $p_1=p_2$. 
	}
\end{figure}

For $p_3 = 1$, the model is symmetric under the exchange $p_1 \leftrightarrow
1-p_2$,  and $0 \leftrightarrow 1$. Thus, along the line $p_2=1-p_1$ the two
absorbing  states 0 and 1 have equal weight, and,  according to
Hinrichsen~\cite{Hinrichsen_1997},  this transition should belong to the PC
universality class. 

\section{Mean-field approximation}
In order to have a qualitative idea of its behavior, we first study our model
within the mean-field approximation. If $c(t)$ denotes the density of occupied
cells at time $t$, we have  
\meq[c]{ 
	c(t+1)&=3p_1c(t)\left(1-c(t)\right)^2+\\
	&\qquad 3p_2c^2(t)\left(1-c(t)\right)+c^3(t).
}
This one-dimensional map has three fixed points:
\eq{
	0,\quad 1, \quad\hbox{and}\quad c^*=\frac{3p_1-1}{1+3p_1-3p_2}.
}
0 is stable if $p_1<\frac{1}{3}$, 1 is stable if $p_2>\frac{2}{3}$, and
$c^*$ is stable if $p_1>\frac{1}{3}$ and  $p_2<\frac{2}{3}$. In the
$(p_1,p_2)$-parameter space, the bifurcations along the lines 
\meq{
	p_1=\frac{1}{3}, 					&		\quad0\le p_2\le\frac{2}{3},\\
	\frac{1}{3}\le p_1\le 1, 	&		\quad p_2=\frac{2}{3},\\
}
may be viewed as second-order transition lines; the first one between the
active and empty states, and the second one between the active and fully
occupied states. 

\begin{figure}[t]
	\begin{center}
	\hspace{0cm}\psfig{figure=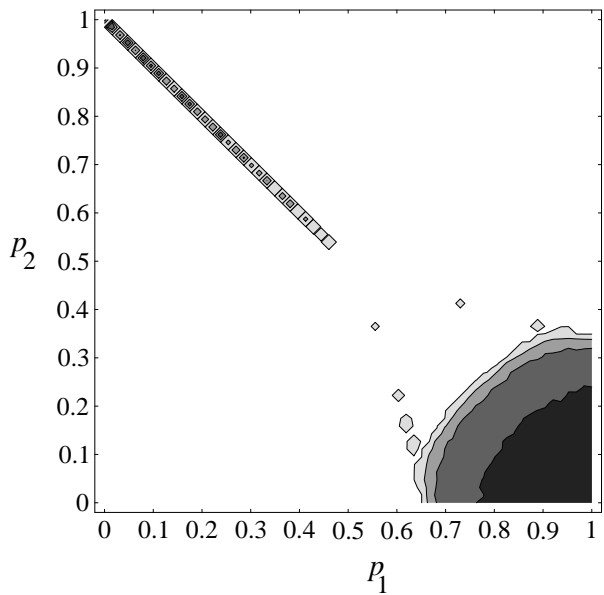,width=8cm}%
	\end{center}
	\caption{Phase diagram for the damage spreading 
		from direct numerical simulations. 
		Numerical values as in Figure 3, color codes as in Figure 2. Traces of the 
		second order phase transitions are present; they join to the first 
		order ($c_0=0.5$) phase boundary. 
	}
\end{figure}

In the domain defined by  
\eq{
0\le p_1<\frac{1}{3}\quad\hbox{and}\quad\frac{2}{3}< p_2\le 1
}
fixed points 0 and 1 are both stable. Their basins of attraction are,
respectively, the semi-open intervals $[0,c^*[$ and $]c^*,1]$. Therefore, if we
start from a uniformly distributed random value of $c$, as time $t$ goes to
infinity, $c(t)$ tends to 0 with probability  $c^*$, and to 1 with probability 
$1-c^*$. Since, for $p_1+p_2=1$, $c^*=\frac{1}{2}$, the line defined by
\eq{
p_1+p_2=1,\quad\hbox{with}\quad 
0\le p_1<\frac{1}{3}\quad\hbox{and}\quad\frac{2}{3}<p_2\le 1
}
is similar to a first-order transition line, although we cannot define a free
energy for  our dynamical system. 

In general, first order phase transitions are associated to the presence of an
hysteresis cycle, due to the coexistence of two phases,  both in equilibrium.
The problem of a proper definition of such a cycle  is rather subtle: since
this model presents absorbing states, it is out of equilibrium and the
ergodicity is always broken, even for finite-size systems.  Thus, once the
system settles into one of the two absorbing states, it never exits,  even if
it is not stable. One can circumvent this problem adding a little of noise,
i.e.\ setting $P(0|000)=1-P(1|111)=\varepsilon$. This assumption, however,
brings the model into the class of equilibrium models, thus forbidding the
presence of a true phase transition in one dimension.  However, if
$\varepsilon$ is small, the system can get trapped for a long time $T$ in a
metastable region near an absorbing state.  The time $T$ is also a function of
the system size $L$. Thus, we have to perform  carefully the limits
$L\rightarrow \infty$, $T\rightarrow\infty$ and $\varepsilon\rightarrow 0$. In
practice, we have observed that the intensity of noise $\varepsilon$ strongly
affects the amplitude of the hysteresis region, while, for a system size $L$
sufficiently large, there exists a large interval of possible time length $T$.
The results from numerical simulations are reported in section 4.

\begin{figure}[t]
	\begin{center}
	\hspace{0cm}\psfig{figure=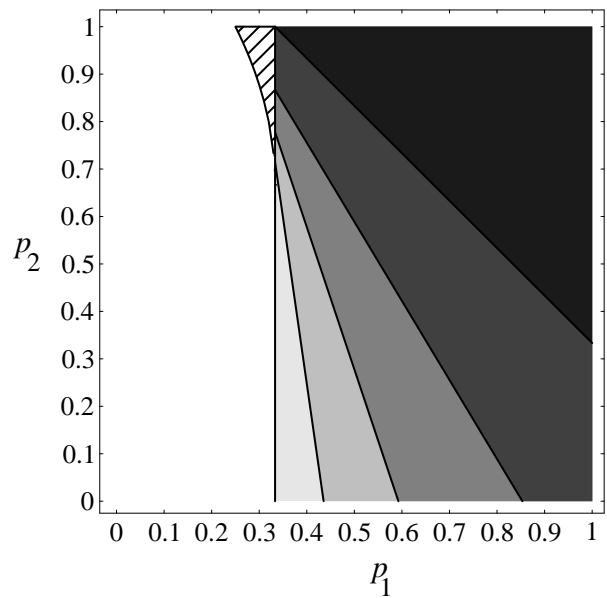,width=8cm}%
	\end{center}
	\caption{
		Mean-field phase diagram for $p_3=0$.  Color codes as in Figure 1. 
	}
\end{figure}

In the mean-field approximation the hysteresis region ranges from $0<p_1<1/3$,
$p_2=2/3$  to $p_1=1/3$, $2/3<p_2<1$.  The mean-field phase diagram for the
density is shown in Figure 1. 

One can write down the mean-field equation for the damage spreading taking into
consideration all possible local configurations of the two lattices.  Let us
denote by $\eta(t)$ the density of damaged sites at time $t$. The evolution
equation for $\eta$ depends on correlations among sites, i.e.\ on the order of
the mean field for the density. Using the simplest factorization for the
density, $\eta(t)$ depends on $c(t)$. The evolution equation for the minimum
damage $\eta$, i.e.\ the damage when the evolution of the two replicas is
computed using only one random  number, is given by 
\meq[eta]{
&\eta(t+1) = 
	 \sum_{\substack{{s_1s_2s_3}\\{h_1h_2h_3}}} 
	\pi(c(t), s_1s_2s_3)
   \pi(\eta(t), h_1h_2h_3)\cdot\\
  &\qquad \big| P(1| s_1s_2s_3) - P(0|
   s_1s_2s_3\oplus h_1h_2h_3)\big|
}
where
\eq{
	\pi(\alpha, x_1x_2x_3) =
	\alpha^{x_1+x_2+x_3}(1-\alpha)^{3-x_1+x_2+x_3}
}
and the symbol $\oplus$ represents the bitwise sum modulus two 
(XOR) of two Boolean
configurations. The value for
$c(t)$ is given by mapping \eqref{c}.     

We have numerically iterated Equation \eqref{eta}.  
The resulting mean-field phase diagram for the damage is shown in
Figure 2. 

\begin{figure}[t]
	\begin{center}
	\hspace{0cm}\psfig{figure=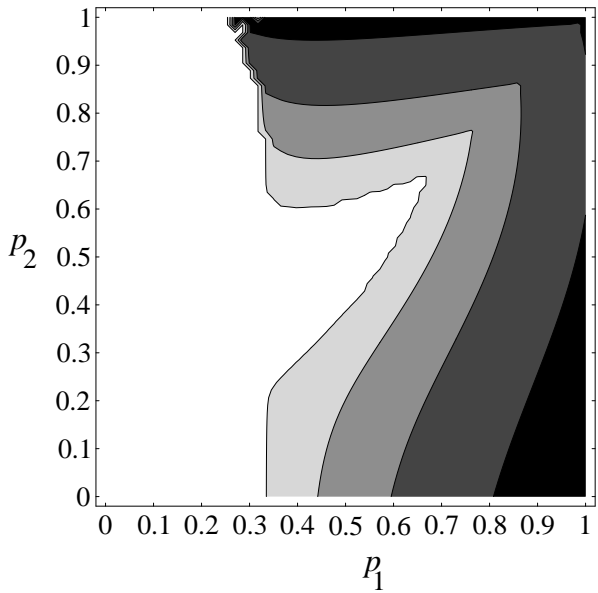,width=8cm}%
	\end{center}
	\caption{Mean-field damage-spreading phase diagram, 
		for $p_3=0$. The diagram has been obtained numerically
		iterating Equation (2).
	}
\end{figure}

\section{Numerical simulations}
We used the fragment method~\cite{Bagnoli_etal_1997} to determine the phase
diagram. Figure 3 represent a plot for the asymptotic density $c$ of occupied
cells when the initial fraction of the occupied cells is equal to $0.5$.  The
scenario is qualitatively the same as predicted by the mean-field analysis. In
the vicinity of the point $(p_1,p_2)=(0,1)$ we observe a discontinuous
transition from $c=0$ to $c=1$, while close to $(p_1,p_2)=(1,0)$ the transition
is continuous. The two second-order phase-transition lines from the non fully
occupied states to either the fully occupied state or the empty state are
symmetric and the critical behavior of the respective order parameters, $1-c$ 
and $c$, are the same. We have checked  that the critical exponent $\beta$ for
these transitions far from the crossing point are numerically the same of DP.

These two transition lines meet on the diagonal $p_1+p_2=1$ and become a
first-order transition line. Crossing the phase boundaries on a line parallel
to the diagonal $p_1=p_2$, the density $c$ exhibits two critical transitions,
as shown in the inset of Figure 3. Approaching the crossing  point, the
critical regions of the two transitions vanishes, the transition itself become
sharper and the corrections to scaling increase.  

\begin{figure}[t]
	\begin{center}
	\hspace{0cm}\psfig{figure=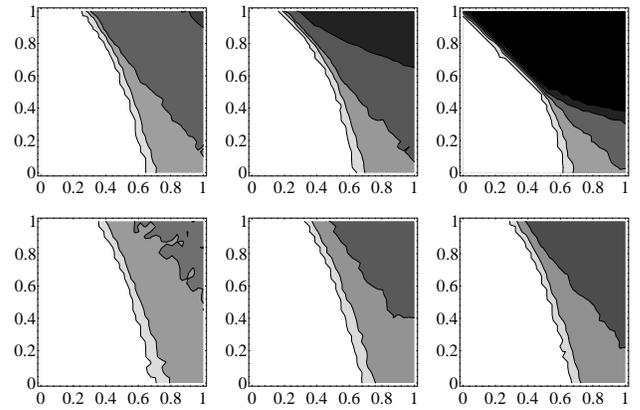,width=8.5cm}%
	\end{center}
	\caption{Cut of the density
		phase space along the $p_1$-$p_2$ plane for different values of $p_3$. From
		left to right and for bottom to top: $p_3=0.0$, $0.2$, $0.4$, $0.6$,
		$0.8$ and $1.0$. Color codes as in Figure~1
	}
\end{figure}

At the crossing point, around $(0.5, 0.5)$, the two attractors have 
symmetrical weight. If we relabel couples of 1's with the symbol 1, couples of
0's with the symbol 0 and couples 01 or 01 with the symbol $A$, we fulfill the
condition stated by Hinrichsen~\cite{Hinrichsen_1997} to 
have symmetric absorbing states, whose
cluster are always separated by an active ($A$) layer. We have performed
preliminary measurements of the exponent $\beta$ for the density of $A$ couples
along the line $p_1+p_2=1$, and obtained $\beta \simeq 0.65(5)$, which  is
consistent with the values for the PC process reported in 
Refs.~\cite{Grassberger_etal_1984,Grassberger_1989}.
  The transition point is at $p_1=0.460(2)$, 
which defines the position of the tricritical point.   
 
In order to study the presence of an hysteresis region, we performed several
scanning of the first-order transition using the local structure
approximation~\cite{Gutowitz_etal_1987}. 
 We cut the phase diagram with a line parallel to the
diagonal $p_1=p_2$,  increasing the value of $p_1$ and $p_2$ after a given
relaxation time up to $p_2=1$; then reverting the scanning up to $p_1=0$.   
The results are reported in Figure 4.  The size of the hysteresis region grows
with the level of the noise  $\varepsilon$. Preliminary simulations show that
the relaxation time $T$ and the noise level $\varepsilon$ scales as
$\varepsilon T^{1.434}= {\rm const}$.

\begin{figure}[t]
	\begin{center}
	\hspace{0cm}\psfig{figure=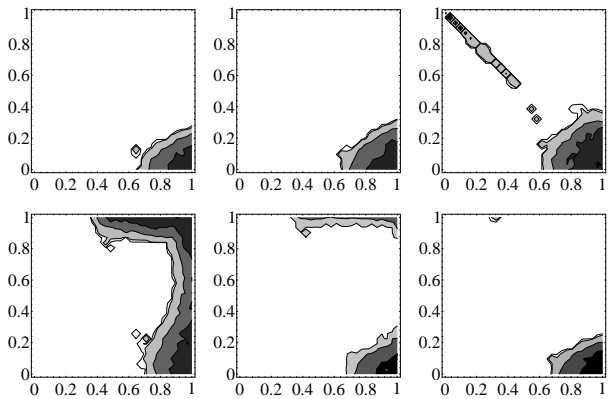,width=8.5cm}%
	\end{center}
	\caption{Cut of the
		damage-spreading  phase space along the $p_1$-$p_2$ plane for several values
		of   $p_3$. From left to right and for bottom to top: $p_3=0.0$, $0.2$,
		$0.4$, $0.6$, $0.8$ and $1.0$. Color codes as in Figure~2.
	}
\end{figure}

The phase diagram for the damage spreading is shown in Figure 5. Outside the
true damage-spreading region, there appear small damaged domains  on the other
phase boundaries. This is due  either to the divergence of the relaxation time
( second-order transitions) or to the fact that a small difference in the
initial configuration can drive the system to a different absorbing state
(first-order transitions).

The ``chaotic'' domain near the point $(p_1,p_2)=(1,0)$ is stable regardless of
the initial density. Note that on the line $p_2=0$ its boundary coincides with
the transition line from the active state to the empty state. Using an argument
similar to the one in Ref.~\cite{Bagnoli_1996} 
would prove that the derivatives of
the two boundary curves coincide, that is, the chaotic phase exhibits a {\it
reentrant} behavior. Here again, the symmetry of the phase diagram implies a
similar behavior on the line $p_1=1$.

\section{Extended phase diagram}

It is interesting to study our model when the transition probability
$P(1|111)=p_3<1$. In this case, there is only one absorbing state. 

\begin{figure}[t]
	\begin{center}
	\hspace{0cm}\psfig{figure=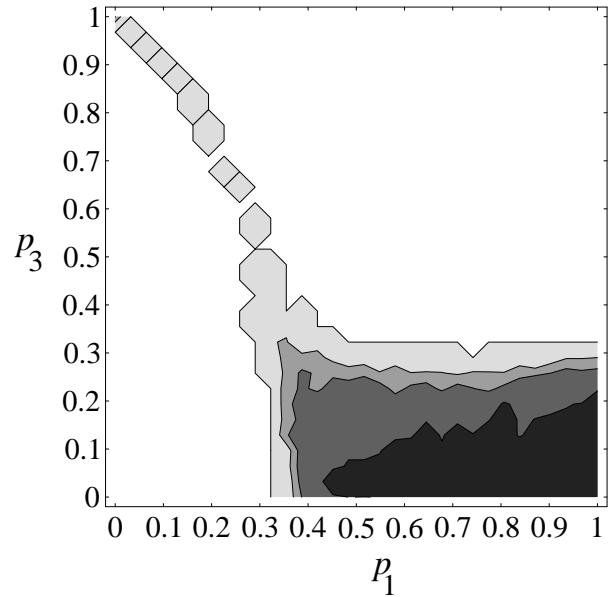,width=8cm}%
	\end{center}
	\caption{Cut of the density phase space along the $p_1$-$p_3$ plane
		for $p_2=1$. Color codes as in Figure 1.
	}
\end{figure}

\subsection{Mean-field approximation}
For some values of $p_3$, the mean-field phase diagram still exhibits a
first-order phase transition. Let us first examine in detail the particular
case $p_3=0$. As for $p_3=1$, the solution $c=0$ looses its stability at
$p=1/3$. The other solutions for the density $c$ are the roots of the quadratic
equation 
\eq{
c^2(p_1-p_2)+c(p_2-2p_1)+p_1-\frac{1}{3} = 0.
}
These two roots are real if $q^2+4(p_1-p_2)/3\ge 0$. This is always the case if
$p_1>1/3$, while for $p_1<1/3$, $p_2$ must lie outside the interval
$\big[\frac{2}{3}(1-\sqrt{1-3p_1}),\frac{2}{3}(1+\sqrt{1-3p_1})\big]$. Since
the density $c$ has to be such that $0\le c\le 1$, only the condition $1\ge
p_2>\frac{2}{3}(1+\sqrt{1-3p_1})$ is meaningful. These solutions are stable.
Therefore, there is a domain in which two stable solutions coexist. The
boundaries of this domain are the three straight lines $p_1=\frac{1}{3}$,
$p_2=1$ and $p_2=\frac{2}{3}(1+\sqrt{1-3p_1})$, where for this last line  $p_2$
belongs to the interval $\big[\frac{1}{4},\frac{1}{3}\big]$. When, inside the
domain,  we approach the point $(p_1,p_2)=\big(\frac{1}{3},\frac{2}{3}\big)$,
the density $c$ tends continuously to zero, which shows that, at this point,
the transition is second-order. The corresponding phase diagram is represented
in Figure 6.

For the chaotic phase, the boundary of the domain of stability is shown in
Figure 7. Close to the point $(p_1,p_2)=(1,0)$ the boundary is almost the same
as for $p_3=1$ but there exists another domain close to the point
$(p_1,p_2)=(1,1)$ which did not exist for $p_3=1$. In the vicinity of $(1,1)$,
very small clusters, which tend to grow, are then fragmented because $p_3=0$.
We have also studied the case $0<p_3<1$. 

\subsection{Numerical simulations}
Our numerical simulations give results in good qualitative agreement with the
mean-field  approximation. Figure 8 shows different phase diagrams in the
$(p_1,p_2)$-plane corresponding to different values of $p_3$. As expected, the
boundary of the active phase ($c\ne 0$) for $p_2=0$ does not vary much  with
$p_3$ since, in this case, the probability to find a cluster of three or more
one is vanishing. No first order phase transition for the density is apparent
by these simulations.

Figure 9 shows the domain of stability of the chaotic phase in the
$(p_1,p_2)$-plane for different values of $p_3$.  The domain of stability of
the chaotic phase in the $(p_1,p_3)$-plane for $p_2=1$ is represented in Figure
10. One can see that there are damage-spreading domains  corresponding to the
density phase boundaries.

\section{Conclusion}
We have studied a radius-1 totalistic PCA whose transition probabilities are
given by $P(1|000)=0$, $P(1|001)=P(1|010)=P(1|100)=p_1$, 
$P(1|011)=P(1|101)=P(1|110)=p_2$, and $P(1|111)=p_3$, within the framework of
the mean-field approximation and using numerical simulations.

For $p_3=1$, the system has two absorbing states and four different phases: an
empty state in which no cell is occupied, a fully occupied state, an active
phase in which only a fraction of the cells are occupied, and a chaotic phase.
This system exhibits:
\begin{enumerate}
	\item A first-order phase transition from the fully occupied state
	to the empty state.
	\item Two second-order phase transitions from the active state to
	the fully occupied state and from the active state to the empty state.
	\item A second-order phase transition from the active state to the
	chaotic state.
\end{enumerate}

For $p_3<1$, our numerical simulations show that there is no first-order
phase transition while the mean-field approximation still predicts the
existence of such a transition.

\end{document}